# Tunable stochasticity in an artificial spin network

Dédalo Sanz-Hernández[*,♦], Maryam Massouras[♦], Nicolas Reyren, Nicolas Rougemaille[3], Vojtěch Schánilec[3,4], Karim Bouzehouane, Michel Hehn, Benjamin Canals[3], Damien Querlioz[5], Julie Grollier, François Montaigne, Daniel Lacour[*]

[*] Corresponding authors: dedalo.sanz@cnrs-thales.fr, lacour8@univ-lorraine.fr

[♦] These authors contributed equally to the work

**Affiliations**

*Unité Mixte de Physique, CNRS, Thales, Université Paris-Saclay, 91767, Palaiseau, France*
Dr. Dédalo Sanz-Hernández, Dr. Nicolas Reyren, Dr. Karim Bouzehouane, Dr. Julie Grollier,

*Université de Lorraine, CNRS, Institut Jean Lamour, F-54000 Nancy, France*
Maryam Massouras, Prof. Michel Hehn, Prof. François Montaigne, Dr. Daniel Lacour

*Université Grenoble Alpes, CNRS, Grenoble INP, Institut NEEL, 38000 Grenoble, France*
Dr. Nicolas Rougemaille, Dr. Benjamin Canals, Vojtěch Schánilec

*Central European Institute of Technology, Brno University of Technology, 61200 Brno, Czech Republic*
Vojtěch Schánilec

*Université Paris-Saclay, CNRS, Centre de Nanosciences et de Nanotechnologies, 91120 Palaiseau, France*
Dr. Damien Querlioz

**Keywords:**

Artificial spin network, stochastic, metamaterial, magnetic domain-wall, computing, Galton board

**Abstract**

Metamaterials present the possibility of artificially generating advanced functionalities through engineering of their internal structure. Artificial spin networks, in which a large number of nanoscale magnetic elements are coupled together, are promising metamaterial candidates that enable the control of collective magnetic behavior through tuning of the local interaction between elements. In this work, the motion of magnetic domain-walls in an artificial spin network leads to a tunable stochastic response of the metamaterial, which can be tailored through an external magnetic field and local lattice modifications. This type of tunable stochastic network produces a controllable random response exploiting intrinsic stochasticity within magnetic domain-wall motion at the nanoscale. An iconic demonstration used to illustrate the control of randomness is the Galton board. In this system, multiple balls fall into an array of pegs to generate a bell-shaped curve that can be modified via the array spacing or the tilt of the board. A nanoscale recreation of this experiment using an artificial spin network is employed to demonstrate tunable stochasticity**.** This type of tunable stochastic network opens new paths towards post-Von Neumann computing architectures such as Bayesian sensing or random neural networks, in which stochasticity is harnessed to efficiently perform complex computational tasks.

**Introduction**

A foreseen stalling of Moore's Law[1] and the success of brain-inspired artificial-intelligence hardware[2–5] have triggered an intense search for spintronic devices that can perform complex computational tasks efficiently.[6] Magnetic vortex nano-oscillators,[4] magnetic domain-wall synapses,[7] and spin-ice networks[8] are part of this emerging family of alternative physics-based computing platforms.[6] In magnetic domain-wall devices, non-volatile information storage, transfer, and processing can occur simultaneously through the controlled movement of oppositely magnetized regions in a magnetic track,[9] placing such devices under intense investigation.[10–13]

Stochasticity in domain-wall motion is highly detrimental for traditional Boolean logic applications, and great efforts have been made to inhibit it in devices.[14,15] However, a powerful set of stochastic computing frameworks exists and is under intense investigation. These frameworks function by generating and transforming probability distributions, and can efficiently deal with data in which uncertainty is intrinsically present, or with difficult optimization tasks.[16–18] In this context, the once detrimental DW stochasticity becomes an exploitable asset, in the form of a natural source of randomness directly deployable within more complex architectures. Here, we demonstrate that the stochastic motion of magnetic domain-walls within an artificial spin network is a controllable and tunable process, by recreating one of the iconic experiments in statistics: the Galton board.[19]

**Main Text**

Our nanoscale magnetic equivalent of a Galton board is composed of an elliptical domain-wall nucleation pad and a bifurcating network of $Ni_{80}Fe_{20}$ magnetic nanowires (Figure 1a), 25 nm thick and 200 nm wide (see Methods). Upon application of an external magnetic field in the x-y plane ($H_1$), the elliptical pad reverses and injects a domain-wall into the network. The domain-wall then travels downwards, taking a random decision at each node it encounters. Magneto-Optical Kerr Effect (MOKE) microscopy (Figure 1b, see Methods) reveals the path taken by the domain-wall. Figure 1c illustrates, for the domain-wall path in Figure 1b, the magnetic orientation of each segment in the board. The '?' symbol indicates the nodes where the domain-wall took a random decision.

In this experiment, the detection of a large number of domain-wall propagation events is required. To achieve this, we optimize a full-field MOKE microscope (see Methods) to obtain 201,852 high-resolution magnetic images, a number hardly reachable by other means such as X-ray imaging or magnetic force microscopy. An automatic extraction of domain-wall paths from this large dataset allows us to precisely probe the properties of each node in the Galton board, the tuneability of the random process, and the degree of randomness in domain-wall path selection.

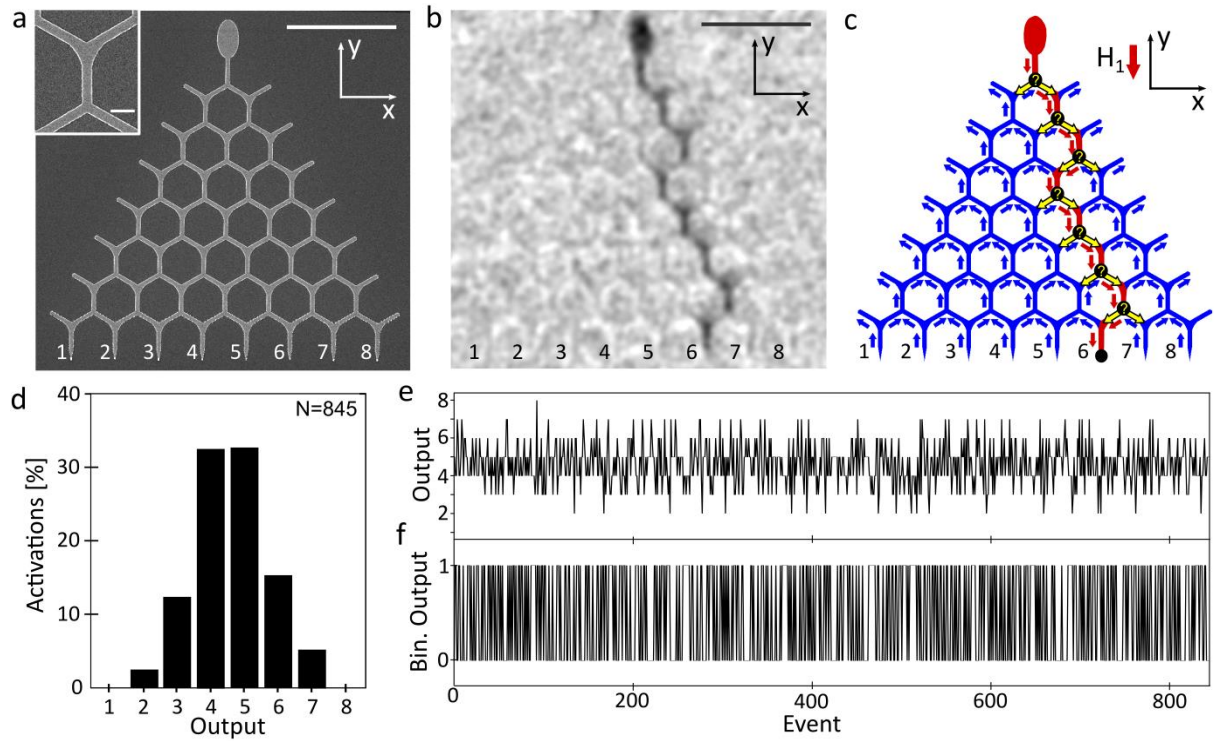

**Figure 1: Nanoscale magnetic Galton board. (a)** Scanning electron micrograph of a nanoscale magnetic Galton board composed by an elliptical nucleation pad at the top and a branching network of magnetic nanowires. Inset: Round edges are employed at nanowire merging points to prevent domain-wall pinning. **(b)** Longitudinal Kerr Effect (MOKE) imaging of magnetic domain-wall propagation. **(c)** Schematics of the magnetization reversal triggered by the domain-wall trajectory in (b). An 18mT uniaxial field ($H_1$) along the -y direction was employed to inject a domain-wall, which took a random decision at each node marked '?'. **(d)** Distribution of activated board outputs after N = 845 single-cascade events in 2,403 field cycles. **(e)** Order in which the 845 output activations occur. **(f)** Binarized output sequence employed to test randomness quality using the NIST test suite, obtained by mapping odd outputs in (e) to 0 and even outputs to 1. Scale bars: (a, b) 5µm, (a, inset) 500nm.

We first study domain-wall motion under symmetric conditions (Field $H_1$ along y), counting the number of times that a domain-wall reaches each of the eight outputs (Figure 1d). Stochastic domain-wall propagation is observed, with the distribution obtained approaching a binomial one. True stochasticity also requires independence between events in the time domain, i.e., the order of output activations (Figure 1e). For a quantitative evaluation of this degree of randomness, we binarized the sequence (mapping even outputs to 1, and odd outputs to 0, see Figure 1f) and passed it through the NIST Statistical Test Suite for Random and Pseudorandom Number Generators,[20] which searches binary sequences for deviations from complete unpredictability. The sequence passes the 13 tests applicable to its length (see Methods), indicating that it comprises a highly uncorrelated set of random events. Another possibility in domain-wall systems is to have short-term memory, leading to correlations between subsequent domain-wall decisions.[21,22] Detailed trajectory analysis in the network reveals that no significant memory is present, which we associate to a chaotic interaction between magnetic vortices and anti-vortices in the domain-wall[23,24] (see Methods).

We continue by investigating the tuneability of the decision process at each node, a critical degree of freedom for the generation of arbitrary distributions. We define '*node probability*' as the probability of a right turn at a given bifurcation, and we evaluate how the node probability depends on the angle θ between the driving field $H_1$ and the y-axis (Figure 2a). Introducing this angle is the magnetic equivalent to misaligning a macroscopic Galton board with respect to the gravitational field.

The structure studied contains 28 nodes, the first six marked '?' in Figure 2a, and as a MOKE image is available for each domain-wall propagation event, node probability can be studied for each node independently, at every field angle. We monitor first the overall evolution of node probability with θ. Figure 2b plots node probability for the 28 nodes as histograms, at θ = 0.0° (green), 3.0° (blue) and -3.0° (orange) respectively. We directly observe that a field tilt of 3.0° leads to a uniform bias in node probability of approximately 25 percentage units. Figure 2c displays the center and standard deviation of the histograms for a larger number of field alignments, revealing that this bias is continuous and smooth in θ. Examining the output probability distributions (Figure 2d), a lateral shift of the distribution is observed as direct consequence of such uniform shift in node probability, analogously to the macroscopic case of a tilted board.

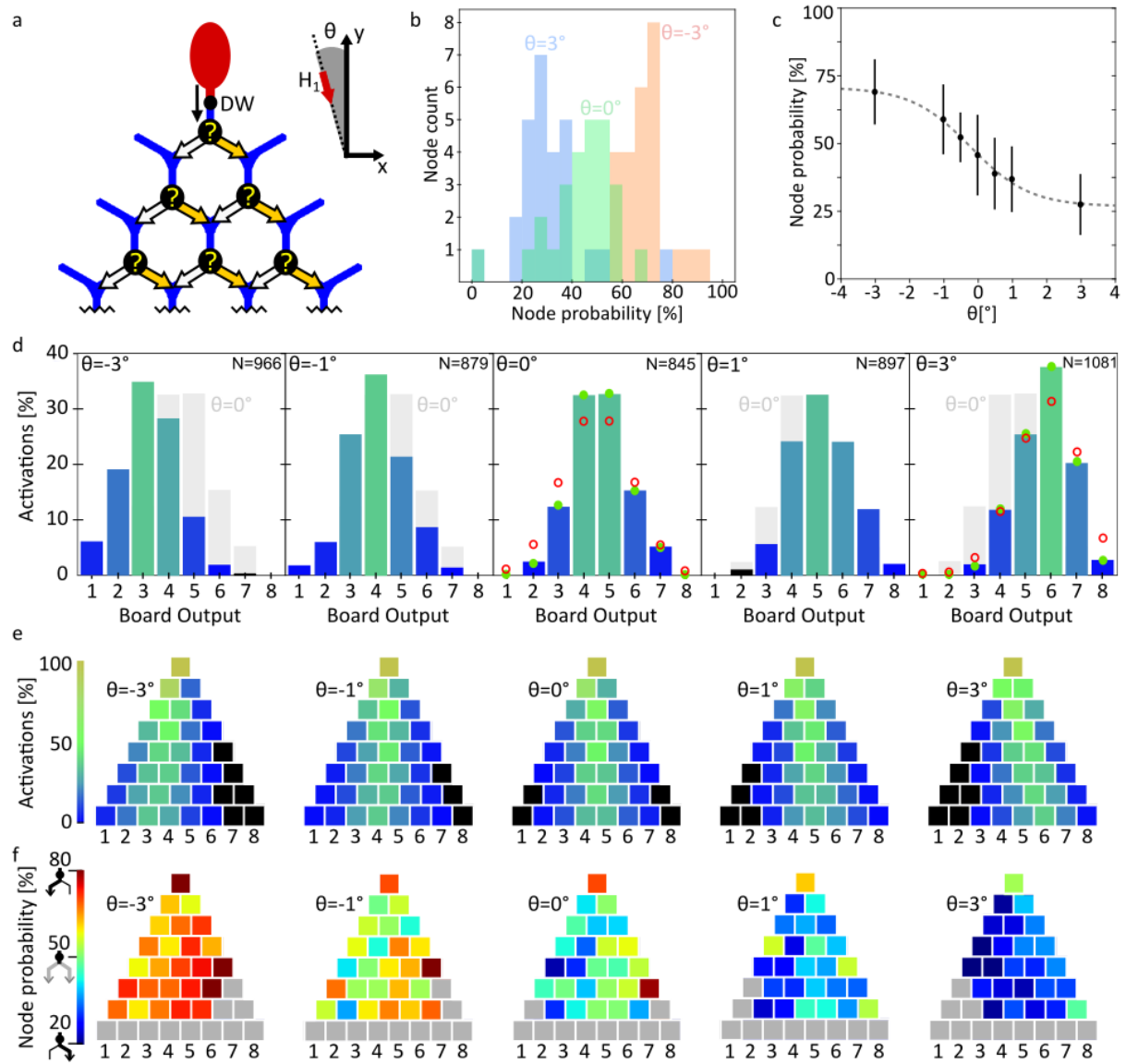

**Figure 2: Tuneability of node probability.** (a) The probability of a magnetic domain-wall (DW) for turning right at each node marked '?' (node probability) is controlled by tuning the angle between the applied field $H_1$ and the symmetry axis of the Galton board (y). A dark-yellow arrow indicates the direction (left turn) favored by the angle θ in the diagram. (b) Histogram of node probability vs. field angle for the 28 nodes in the Galton board studied in Figure 1. (c) Mean and standard deviation of the distribution in node probability as a function of field angle θ. Dashed line: guide to the eye hyperbolic tangent fit. (d) Effect of field alignment on the magnetic Galton board output distribution. In color: output distribution, light-grey shade: distribution at θ=0.0°. Color follows the scale in (e). N indicates the number of events in each distribution after 2,403 field cycles. Red circles and green dots in panels θ=0.0° and θ=3.0° indicate the theoretically expected distribution for an ideal Galton board with the same probability in all nodes (red) and for a board in which each probability corresponds to the measured probability in panel (f) (green). **(e)** Spatial distribution of node activation as a function of θ. **(f)** Spatial distribution of node probability values. Confidence intervals given in Supplementary Figure 3. Grey nodes indicate impossibility to perform the measurement (insufficient counts/not-applicable). **(b-f)** ±0.3° experimental uncertainty in the alignment between the θ=0.0° direction and the y axis.

To confirm the behavior, we further characterize the complete network (Figures 2e-f), looking at the number of trajectories that went through each node (Figure 2e) and evaluating locally all node probabilities (Figure 2f) by measuring the fraction of domain-walls that took a right turn. The expected tilting in the distribution of domain-wall trajectories is observed in Figure 2e, and a narrow distribution in node probability is observed at all fields (See also Figures 1b-c) with no obvious spatial correlations, indicating that variability between nodes is likely caused by small random differences during nanofabrication.

The red circles in Figure 2d show the expected output distribution of an ideal Galton board network with uniform probabilities of 50% (θ = 0.0°) and 68% (θ = 3.0°) respectively. These red circles do not coincide with the observed distributions. However, if we calculate numerically the distribution resulting from cascading the individual node probabilities measured experimentally in Figure 2f, an accurate match is observed, represented by green dots in Figure 2d. This shows that the generated output distribution can be precisely computed by knowing the individual node probabilities, and can be controlled through these probabilities.

Results in Figure 2 set the foundations for creating networks in which node probability is tuned locally by means of electrical currents or geometrical modifications, to generate arbitrary output distributions. Given the wide and continuous tuneability range of node probabilities demonstrated (Figure 2e), generating an arbitrary probability distribution would only require a simple numerical optimization protocol.

Finally, we illustrate how the lattice structure can be engineered to further tune the output distribution and its normalization. In a complete lattice, all possible paths through the structure lead to an output, and therefore a consistent number of activations (N) is obtained at all operation angles. The structure in Figures 3a and b, however, breaks this condition by removing the central vertical element and leaving the oblique segments marked *. These elements act as domain-wall sinks, preventing them from reaching an output and changing the resulting distribution. These sinks are the nanoscale equivalent to drilling a hole at the back of a macroscopic Galton board.

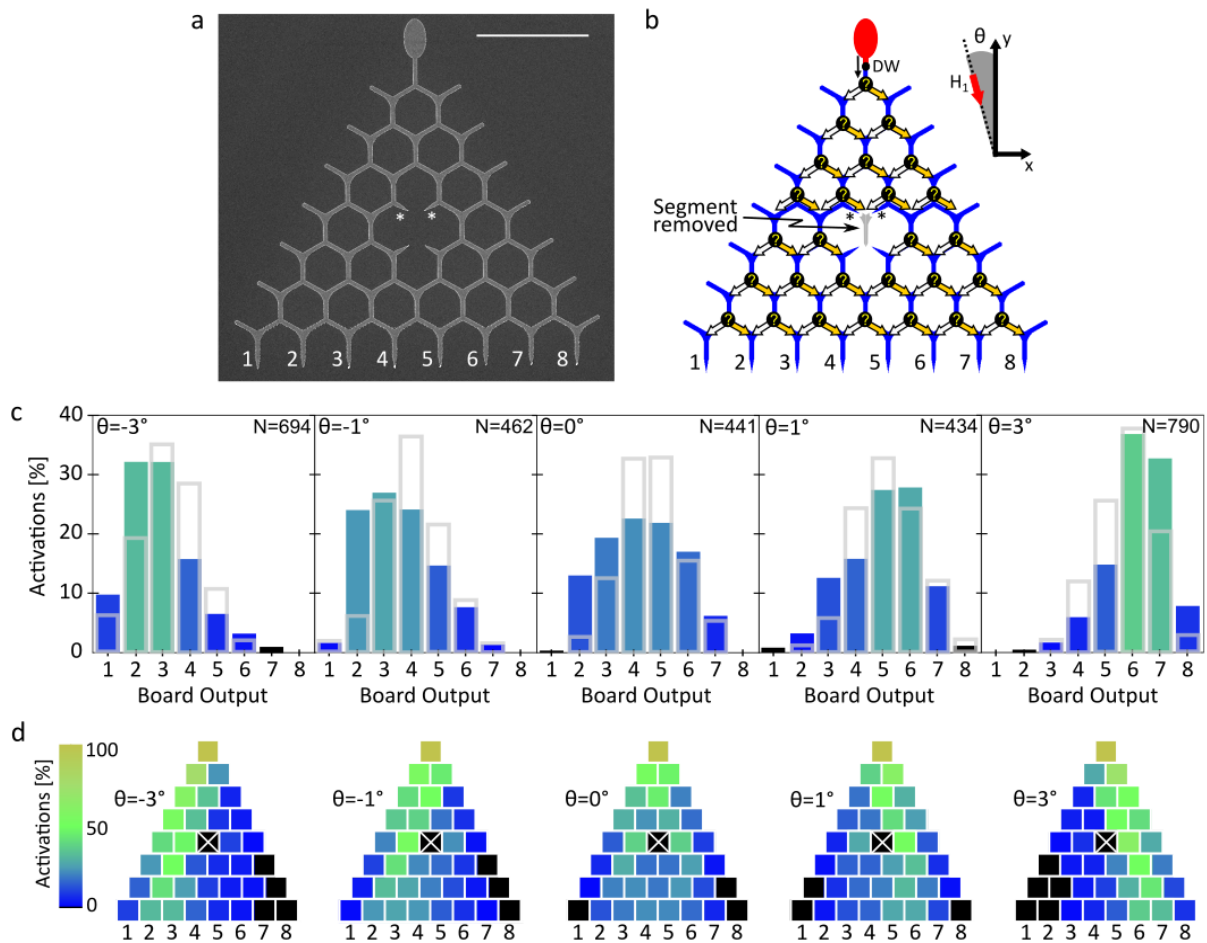

**Figure 3: Galton board lattice edition.** (**a**) Scanning electron micrograph of a nanomagnetic Galton board with a missing central element. The segments marked '*' act as domain-wall sinks. (**b**) The probability of a magnetic domain-wall (DW) for turning right at each node marked '?' (node probability) is controlled by tuning the angle between the applied field $H_1$ and the symmetry axis of the Galton board (y). (**c**) Effect of field alignment θ on the distribution of output activations. Grey lines display the distribution of a complete board. N indicates the number of events in the distribution after 2,403 field cycles. (**d**) Spatial distribution of node activation as a function of θ. The node removed is indicated with a white cross. Scale bar: 5µm.

The measured consequences in the output distributions are presented in Figures 3c and d. At θ = 0.0°, a significant flattening of the distribution occurs compared to the complete network (see solid bars vs. grey lines), as well as a reduction of the number of detected events N. For tilted fields, the lateral shift of the output distribution is larger than in the complete board, while the reduction in N is less pronounced. These effects can be directly correlated to the number of domain-walls that travel through the central segment in Figure 2e, i.e., a reduction of 50% in N is observed at θ = 0.0°, whereas a reduction of 30% in N is observed at θ = ±3.0°. In future devices, strategic positioning of domain-wall sinks or stochastic pinning traps may be exploited in addition to node probability tuning to control the output distributions.

**Conclusion**

In conclusion, the suitability of artificial spin networks to generate and sample tunable probability distributions is demonstrated by evaluating key randomness metrics in a nanoscale magnetic Galton board. Continuous tuneability of the domain-wall path-selection process is achieved and highly-uncorrelated randomness is demonstrated in the time and space domains. Further tuneability is achieved by using domain-wall sinks, which selectively destroy domain-walls passing through some nodes. These results set the ground for the design of artificial spin networks with an arbitrary stochastic response to the injection of magnetic information.

**Methods**

**Nanostructure fabrication:** The artificial spin network is composed by a 25 nm thick, 200 nm wide Permalloy track and curved edges are employed when nanowires merge into vertical segments (Figure 1a, inset) to increase the number of single-cascade events. The device was patterned using electron-beam lithography. After resist development, 25 nm of Permalloy (Atomic fractions: Fe 20.49% and Ni 79.51%) were thermally evaporated and capped with 3nm of Aluminum. Resist lift-off was performed under gentle sonication.

**MOKE detection:** For 2,403 field cycles of 12 steps, at 7 field orientations (201,852 measurements in total), the paths taken by domain-walls were imaged using a Magneto-Optical Kerr Effect (MOKE) microscope (EVICO Magnetics GmbH) equipped with a 100x lens with 1.3 numerical aperture (Zeiss Epiplan-NEOFLUAR) in immersion oil (n = 1.518, Immersol$^{TM}$ M). Images are acquired using a "longitudinal mode" with a static in-plane (y-axis) component of the light momentum. The image sequences are automatically extracted using software. 7 images are taken and averaged for each measurement in order to improve signal to noise ratio (1,412,964 images were processed in total). Each compound image was compared to a saturated state to extract switching events. The dataset was acquired continuously over the course of 10 days, occupying 4Tb of disk space. To have an accurate estimate of the node probability, only the cascades in which one single element switches for every row of the board are considered for analysis. Depending on operating conditions, the fraction of single-cascades detected ranges between 38% and 51%, with most of the remaining fraction corresponding to incomplete events in which the domain-wall does not reach the bottom of the board. An improvement of this fraction is expected upon further optimization of the fabrication process. When applicable, the label N in all Figures indicates the number of detected events. All images were acquired using the same (static) off-axis illumination conditions.

**Field operation:** Magnetic fields were applied using a computer-controlled quadrupole electromagnet coplanar to the device. An experimental uncertainty of ±0.3° exists in the alignment between the quadrupole's y-axis and the board's symmetry axis.

An operating field of 18 mT was chosen to maximize the number of single-cascade events, in which one single element switches for every row of the board (see Supplementary Figure 1a). We observe that domain-wall injection from the pad into the top input node occurs consistently at fields between 7 and 10 mT (red in supplementary Figure 1-b). 25% of these domain-walls travel to the second layer of the device within the same field range and the rest require a larger field value which is between 16 and 18 mT (blue in supplementary Figure 1-b). No output activations are detected at low fields, with most events requiring between 16 and 18 mT (green in supplementary Figure 1-b).

We identify a significant number of low-field domain-wall injections into the first layer (See blue histogram in Supplementary Figure 1-b) as the most-likely cause of the bias in node probability observed for the top node in the structure (see Methods: Further node characterization). Exploration of the exact nature of this process is left for future studies.

**NIST Statistical Test Suite:** The NIST SP 800-22 Statistical Test Suite[20] was developed by NIST (US National Institute of Standards and Technology) as a set of 15 tests for detecting deviations of binary sequences from randomness, with a special focus on testing random and pseudorandom number generators in the context of cryptographic key generation. We employ a pre-existing Python implementation[25] to test the binarized sequence produced by the nanoscale Galton board.

Each test in the suite checks for a particular type of deviation from randomness, generating a p-value or 'tail probability' in the range 0 to 1, which describes the probability of a truly random sequence giving an equal or worse test metric than the sequence provided. After applying each test to a sequence, a conclusion is derived that accepts or rejects whether the generator is producing random values if the p-value is above an arbitrary threshold. A 1% threshold is used as a standard. In our study, two changes with respect to the default values were made in order to accommodate for the size of the available dataset (845 events): binary matrix rank test size was reduced to 12 by 12, and the pattern size for the overlapping template matching test was reduced to 4. The Maurer's "Universal Statistical" Test and the Linear Complexity Test have very long string length requirements ($>10^5$ and $>10^6$ respectively) and therefore could not be applied to the available sequence of 845 events. This does not mean that the tests were failed, only that the sequence could not be tested for these particular types of deviation from randomness.

The random binary string of bits obtained from the magnetic Galton board passes all of the applicable tests in the suite with consistently large p-values, indicating that the sequence does not differ significantly from a random string of bits. Supplementary Table 1 displays the results of the 13 tests performed.

**Further node characterization:** Supplementary Figures 2 and 3 show more detailed data regarding the dependence between node probability and field angle for each individual node. In supplementary Figure 2 we observe a singular behavior in the injection node (top of the board), which consistently shows node probability values 20 to 30 percentage units higher than the rest of nodes. We correlate this significant bias with the presence of low-field domain-wall injection events into the first node (see Methods: Field operation).

Supplementary Figure 3 shows the 75% confidence interval in the determination of node probability for each node in the Galton board, this confidence interval is different for each node as a different number of domain-walls pass through each. Confidence intervals were calculated using the Python function: *statsmodels.stats.proportion.proportion_confint()*.

Supplementary Figure 4 quantifies the degree of correlation between domain-wall decisions at subsequent nodes. To quantify this correlation, we model node probability as the combination of an external bias $\eta$ and a "memory bias" $\delta$ related to the previous domain-wall decision. Node probability $P$ (the fraction of domain-walls that turn right at a given intersection), can then be expressed as: $P = \frac{1}{2} + \eta \pm \delta$, where the sign before $\delta$ depends on whether the domain-wall took a right (+) or left (-) turn to reach the node. Positive (negative) values of $\delta$ would therefore favor equal (opposite) decisions between consecutive nodes, respectively leading to zig-zag (armchair) trajectories. To evaluate $\eta$ and $\delta$, we define 4 types of trajectory that can go through a given node (Supplementary Figure 4a) and count the number $z_1$, $z_2$, $a_1$, $a_2$ of each of these trajectories going through each node. We then evaluate the ratios defined by:

$$x_1 = \frac{a_1}{z_1+a_1} = 1 - \frac{z_1}{z_1+a_1} = \frac{1}{2} + \eta - \delta \qquad (1)$$

$$x_2 = \frac{z_2}{z_2+a_2} = 1 - \frac{a_2}{z_2+a_2} = \frac{1}{2} + \eta + \delta \qquad (2)$$

to obtain $\eta$ as $(x_1 + x_2 - 1)/2$ and $\delta$ as $(x_2 - x_1)/2$.

From the results displayed in Supplementary Figures 4b-c, we observe that the deviations of node probability from the 50% equilibrium are strongly determined by $\eta$ (external bias) and only a constant residual contribution of $\delta$ (domain-wall memory) is present. These two metrics have only been evaluated for non-edge nodes, as they are not defined at the edge.

**Micromagnetic simulations:** To further understand the complex domain-wall dynamics leading to a stochastic path selection, we have performed dynamic micromagnetic simulations using version 3.9.3 of Mumax3[26] in an Nvidia GeForce RTX 2080-Ti graphics processor. We perform simulations within a unit cell of the Kagome nanowire lattice (Supplementary Figure 5), with the following parameters: track width 200nm, sample thickness 25nm, Msat = 800e3 [A/m], Aex = 13e-12 [J/m], alpha = 0.02.

During these simulations, we observe complex transformations of the domain-wall, characterized by the repeated injection and ejection of vortex and anti-vortex structures through the track boundaries. This effect has been previously associated to the onset of stochastic path selection in similarly-sized systems[23] and we observe that complex chaotic interactions repeatedly take place between the main vortex domain-wall and the injected anti-vortex structures. Supplementary Figure 6 describes an example of the types of micromagnetic structures observed in these simulations, which are provided in full as Supplementary Video 1. We also observe that the most likely cause of unsuccessful domain-wall propagation is the annihilation of all vortex and antivortex structures upon domain-wall arrival to the curved edges (See Supplementary Video 1). This may be inhibited in the future by further optimization of the geometry. We observe an 80.4% fraction of successful cascades in the simulation of a single unit cell. This rate is consistent with the observed success rate of 30 to 50% observed experimentally for a complete board, in which 6 propagations through a unit cell need to be successful in sequence ($0.804^6 = 0.27$).

We also observe that the direction of domain-wall propagation at an intersection is likely determined by the nucleation and propagation of an edge magnetic defect at one side of the junction (See white arrows in Supplementary Figure 7).

The results of these simulations are in agreement with existing works on stochastic domain-wall dynamics,[14,15,27–29] stochastic domain-wall pinning[14,30,31] and depinning,[32,33] and in which inhibition of Walker breakdown has been shown to eliminate stochasticity.[14,15]

Supplementary Video 1 technical details: This video was created by concatenating simulation frames taken every 0.1 ns. The frame number is indicated at the top-left corner of the video. A new system initialization is forced every 21ns to provide statistics. Residual torques in the dynamic solver of version 3.9.3 of Mumax3 provided the slightly different initial conditions necessary to obtain different domain-wall propagation paths within the same simulation.

**Acknowledgements**

We acknowledge financial support from the ANR (ANR-17-CE24-0007), the Region Grand Est through its FRCR call (NanoTeraHertz and RaNGE projects), by the impact project LUE-N4S part of the French PIA project "Lorraine Université d'Excellence", reference ANR-15IDEX-04-LUE and by the "FEDER-FSE Lorraine et Massif Vosges 2014-2020", a European Union Program. D.S.H. and J.G. acknowledge support by DOE BES Award # DE-SC0019273 (Kerr microscopy, numerical data analysis and manuscript writing) for Q-MEEN-C, an Energy

Frontier Research Center funded by the U.S. Department of Energy (DOE), Office of Science, Basic Energy Sciences (BES). We thank I. Soldatov and R. Schäffer (Evico Magnetics GmbH) for their support with Kerr microscope maintenance and software customization.

**Author contributions**

JG, FM, MM, MH and DL conceived the study. MM and FM fabricated the samples assisted by MH and DL. MM, FM, MH and DL optimized sample design using magnetic force microscopy experiments by MM and KB. FM and DSH performed exploratory micromagnetic simulations. VS and NRo performed exploratory MFM measurements. DSH designed, optimized and performed the Kerr experiments assisted by NRe. DSH performed the data analysis. DSH wrote the manuscript with input from all authors.

**Financial Interests.** The authors declare no competing financial interests.

Supplementary Material

Tunable stochasticity in an artificial spin network

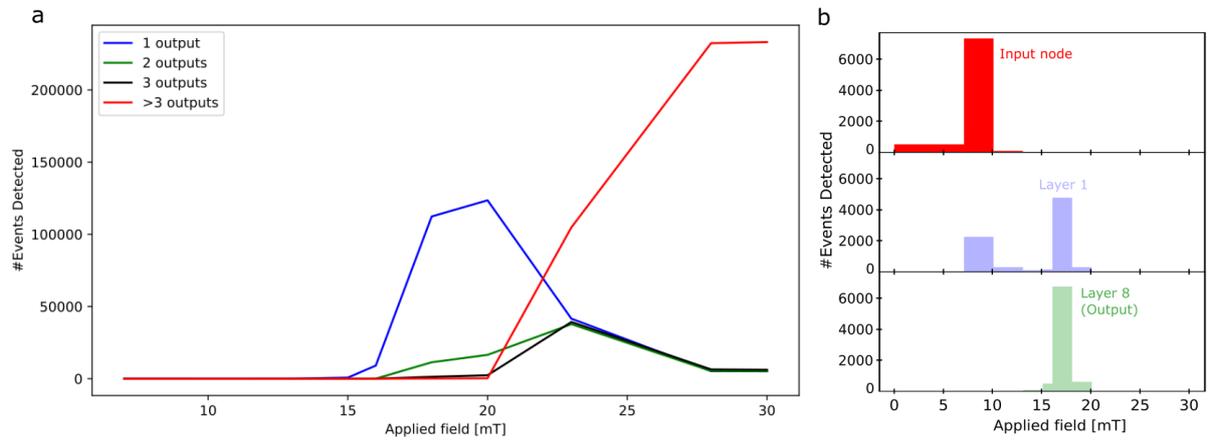

**Supplementary Figure 1. (a)** Dependence of the number of triggered cascades (number of active node outputs at the end of a board) as a function of the applied magnetic field. At 18mT the largest fraction of single cascades is observed. This data is a collection over 201,852 parallel measurements in 16 Galton boards. **(b)** Single-cascade domain-wall propagation fields at different layers (Input, Layer 1, and Output) of the board studied in the main manuscript. Domain-wall injection into the first node occurs at fields between 7 and 10 mT, whereas output activations occur between 16 and 18mT. A fraction of domain-walls propagates into the first layer at fields between 7 and 10 mT, but most events require between 16 and 18mT to reach the first layer.

| NIST 800-22-1a results | | | | |
|---|---|---|---|---|
| | (P-Value, Conclusion) | | | |
| 2.01. Frequency Test: | (0.9725573495328714, True) | | | |
| 2.02. Block Frequency Test: | (0.9108775391455062, True) | | | |
| 2.03. Run Test: | (0.4700094445597227, True) | | | |
| 2.04. Run Test (Longest Run of Ones): | (0.23390365224944776, True) | | | |
| 2.05. Binary Matrix Rank Test: | (0.36113614859914944, True) | | | |
| 2.06. Discrete Fourier Transform (Spectral) Test: | (0.9057648942668173, True) | | | |
| 2.07. Non-overlapping Template Matching Test: | (0.8012941506004918, True) | | | |
| 2.08. Overlapping Template Matching Test: | (0.9999974195044041, True) | | | |
| 2.09. Universal Statistical Test: | Insufficient string length | | | |
| 2.10. Linear Complexity Test: | Insufficient string length | | | |
| 2.11. Serial Test: | (0.6181376590294971, True) | | | |
| | (0.6647426596797406, True) | | | |
| 2.12. Approximate Entropy Test: | (0.9999999710346387, True) | | | |
| 2.13. Cumulative Sums (Forward): | (0.9654594835576138, True) | | | |
| 2.13. Cumulative Sums (Backward): | (0.9783036485547041, True) | | | |
| 2.14. Random Excursion Test: | STATE | xObs | P-Value | Conclusion |
| | '-4' | 10.324647338971081 | 0.06654206483435049 | True |
| | '-3' | 9.551443478260866 | 0.08898953629858565 | True |
| | '-2' | 2.4380032206119164 | 0.785801580136233 | True |
| | '-1' | 1.2608695652173911 | 0.9389061922784552 | True |
| | '+1' | 2.217391304347826 | 0.8183199664928124 | True |
| | '+2' | 2.1878690284487385 | 0.8225862685153571 | True |
| | '+3' | 0.85446956521739140 | 0.9734298060786183 | True |
| | '+4' | 6.319069952737084 | 0.276398147635294 | True |
| 2.15. Random Excursion Variant Test: | STATE | COUNTS | P-Value | Conclusion |
| | '-9.0' | 23 | 0.5608496189974651 | True |
| | '-8.0' | 22 | 0.5182416111597716 | True |
| | '-7.0' | 24 | 0.5246809886375939 | True |
| | '-6.0' | 21 | 0.4319446671778435 | True |
| | '-5.0' | 22 | 0.404248494739471 | True |
| | '-4.0' | 31 | 0.5544649406427751 | True |
| | '-3.0' | 42 | 0.8520519248014127 | True |
| | '-2.0' | 48 | 0.9041774974987763 | True |
| | '-1.0' | 50 | 0.6766573217164245 | True |
| | '+1.0' | 40 | 0.5316145768816123 | True |
| | '+2.0' | 47 | 0.9520019793025064 | True |
| | '+3.0' | 52 | 0.7796682079857953 | True |
| | '+4.0' | 57 | 0.6646800940267454 | True |
| | '+5.0' | 66 | 0.4870251915949487 | True |
| | '+6.0' | 57 | 0.729506525490593 | True |
| | '+7.0' | 35 | 0.7504294025292763 | True |
| | '+8.0' | 17 | 0.4350061647690232 | True |
| | '+9.0' | 13 | 0.4040325964891788 | True |

**Supplementary Table 1.** P-value results of running the sequence in Figure 1-f through the NIST SP 800-22 test suite. The binary string passes all of the applicable tests in the suite. Tests 2.09 and 2.10 are not applicable due to their string length requirements.

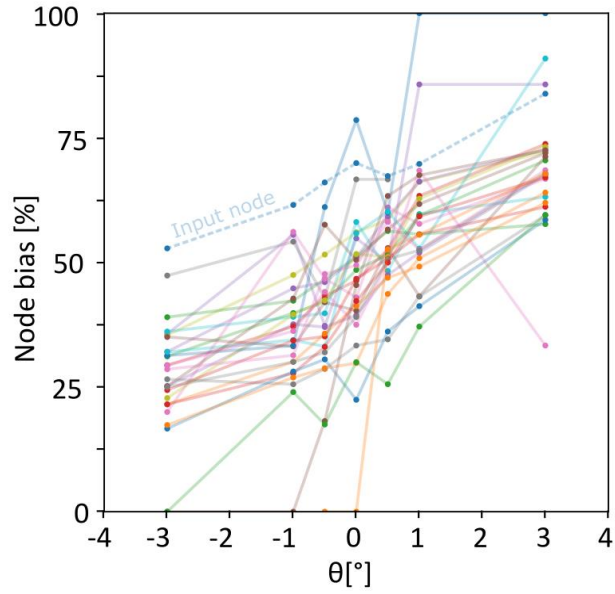

**Supplementary Figure 2.** Dependence of the node probability for the 28 nodes in the structure studied in Figure 2. Each line corresponds to a node, with the injection node being indicated by a dashed line.

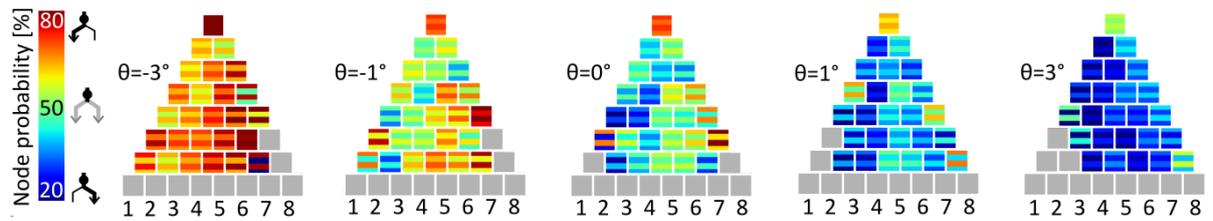

**Supplementary Figure 3.** 75% confidence intervals for the Measurements of node probability presented in Figure 2 of the main manuscript. In this figure each node contains 4 lines with 2 alternating colors that correspond to the 75% confidence interval of the measurement (a different number of domain-walls travel through each part of the board, making the width of this interval vary).

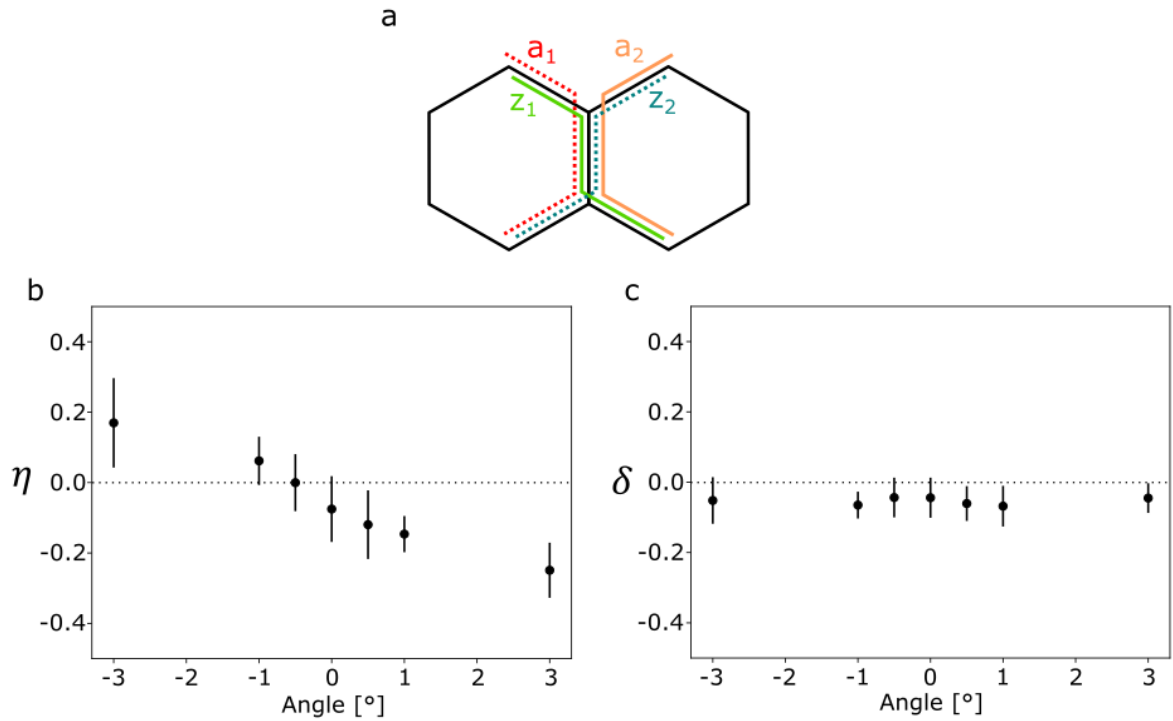

**Supplementary Figure 4.** Analysis of the deviation of node probability with respect to the 50% equilibrium. **(a)** definition of armchair ($a_1$, $a_2$) and zigzag ($z_1$, $z_2$) trajectories used to discern between bias and memory. **(b)** Angular dependence of $\eta$, the external bias contribution to deviations of node probability from 50%. **(c)** Angular dependence of $\delta$, the domain-wall memory contribution to deviations of node probability from 50%. Dots indicate mean values of $\eta$ and $\delta$ found across the 15 non-edge nodes in the structure, error bars indicate standard deviation in the values found.

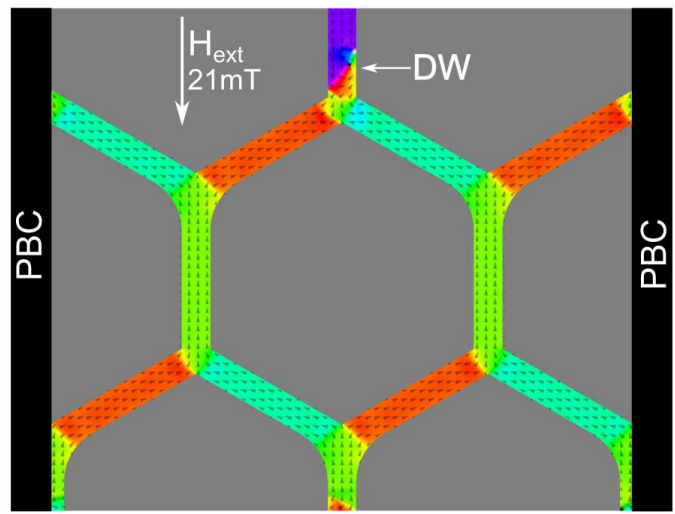

**Supplementary Figure 5.** Simulated unit cell for evaluation of domain wall (DW) trajectories. Periodic boundary conditions, marked PBC, are employed to reduce the simulation volume.

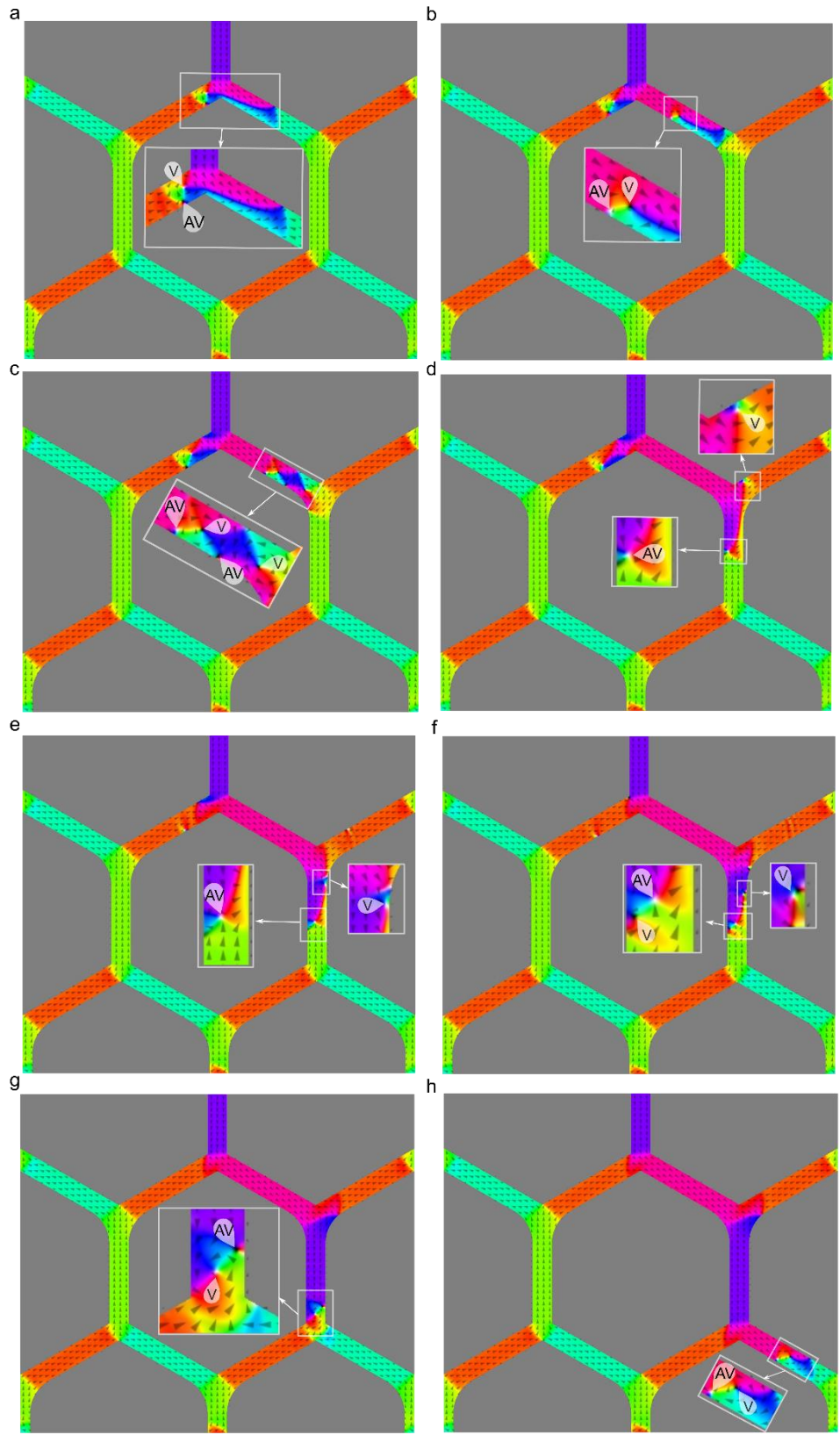

**Supplementary Figure 6.** Simulation snapshots showing propagation of magnetic vortices (marked V) and anti-vortices (marked AV) through the nanowire network. Snapshots extracted from Supplementary Video 1, in order (a-h)

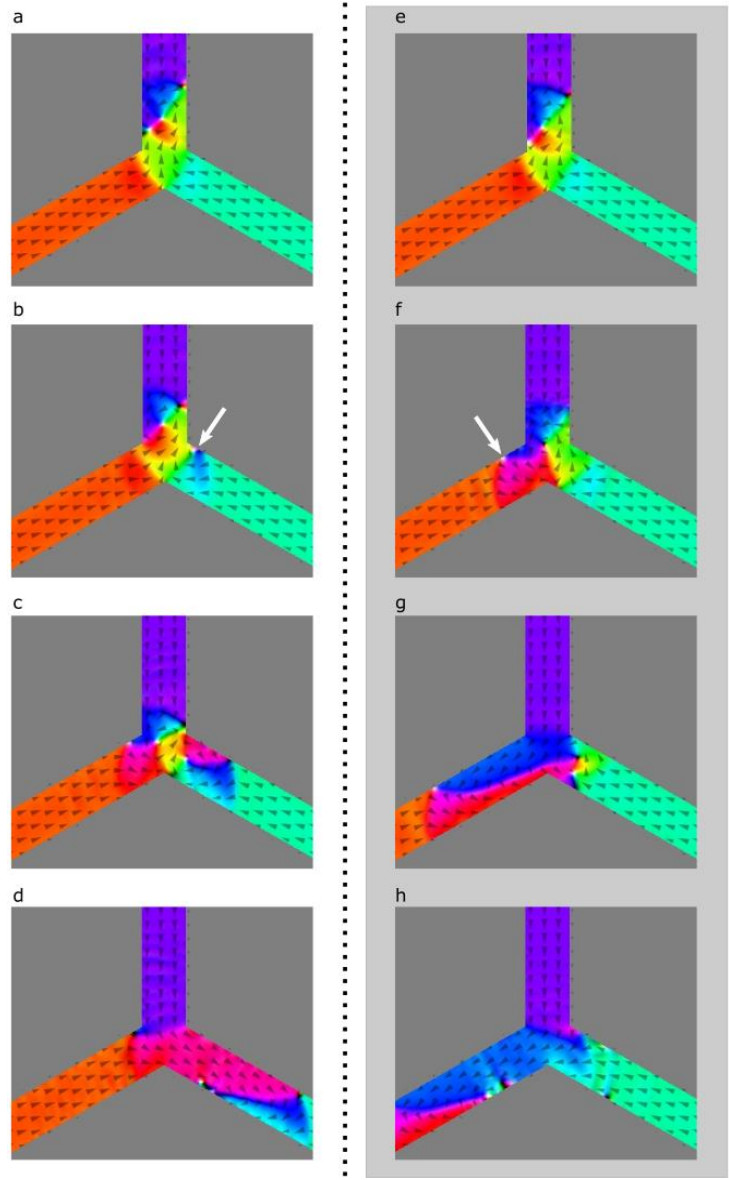

**Supplementary Figure 7.** Comparison of simulations for left (a-d) and right (e-h) domain-wall turns. We identify the nucleation of edge magnetic defects (white arrows) as the most likely driver in the decision-making process. **(a-d)** Simulation snapshots for a left turn. **(e-h)** Simulation snapshots for a right turn.